\documentclass[12pt,preprint]{aastex}

\usepackage{emulateapj5,psfig}
\usepackage{psfig}

\newenvironment{inlinefigure}{
\def\@captype{figure}
\noindent\begin{minipage}{0.999\linewidth}\begin{center}\small}
{\end{center}\end{minipage}\smallskip}

\def\gsim{~\rlap{$>$}{\lower 1.0ex\hbox{$\sim$}}}
\def\lsim{~\rlap{$<$}{\lower 1.0ex\hbox{$\sim$}}}

\def\kms{\,\hbox{km}\,\hbox{s}^{-1}}

\def\Msol{\mathrel{\rm M_{\odot}}}

\begin{document}

\title{GMOS Integral Field Spectroscopy of a merging system with
  enhanced Balmer absorption}

\lefthead{GMOS-IFU Spectroscopy of a Merging System with Enhanced
  Balmer Absorption}
\righthead{Swinbank et al.}

\author{
A.\,M.\ Swinbank,\altaffilmark{1}
M.\,L.\ Balogh,\altaffilmark{1,2}
R.\,G.\ Bower,\altaffilmark{1}
G.\,K.\,T.\, Hau, \altaffilmark{1}
J.~R.\, Allington Smith, \altaffilmark{1}
R.\,C.\ Nichol, \altaffilmark{3,4}
C.\,J.\ Miller \altaffilmark{4}
}

\setcounter{footnote}{0}

\altaffiltext{1}{Department of Physics, University of Durham, South
  Road, Durham DH1 3LE, UK -- Email: a.m.swinbank@dur.ac.uk}
\altaffiltext{2}{Department of Physics, University of Waterloo,
  Waterloo, ON, Canada N2L 3G1} \altaffiltext{3}{Institute for
  Cosmology and Gravitation, Mercantile House, Hampshire Terrace,
  University of Portsmouth, Portsmouth, UK PO1 2EG}
\altaffiltext{4}{Department of Physics, Carnegie Mellon University,
  5000 Forbes Avenue, Pittsburgh, PA 15213}

\setcounter{footnote}{0}

\begin{abstract}
  In this paper we present the three dimensional dynamics of the galaxy
  SDSS\,J101345.39+011613.66, selected for its unusually strong Balmer
  absorption lines (W$_{\rm o}$(H$\delta$)=7.5\AA).  Using the
  GMOS-South IFU in Nod \& Shuffle mode we have mapped the continuum
  and optical absorption lines of this $z$=0.1055 field galaxy.  This
  galaxy has a disturbed morphology, with a halo of diffuse material
  distributed asymmetrically toward the north.  Using the [O{\sc ii}]
  emission line (W$_{\rm o}$[O{\sc ii}]=4.1\AA) we find that the gas
  and hot OB stars are offset from the older stars in the system.  The
  gas also has a spatially extended and elongated morphology with a
  velocity gradient of $100\pm20\kms$ across 6\,kpc in projection.
  Using the strong $H\gamma$ and $H\delta$ absorption lines we find
  that the A- stars are widely distributed across the system and are
  not centrally concentrated arguing that the A-star population has
  formed in molecular clouds outside the nucleus.  By cross correlating
  the spectra from the datacube with an A-star template we find
  evidence that the A-star population has a 40$\kms$ shear in the same
  direction as the gas.  The disturbed morphology, strong colour
  gradients and strong H$\delta$ and H$\gamma$ absorption lines in
  SDSSJ\,101345.39 argue that this is a recent tidal interaction/merger
  between a passive elliptical and star-forming galaxy.  Although based
  on a single object, these results show that we can spatially resolve
  and constrain the dynamics of this short lived (yet important) phase
  of galaxy formation in which the evolutionary process take galaxies
  from star-forming to their quiescent end products.
\end{abstract}


\section{Introduction}
Galaxies with strong Balmer absorption lines in their spectra
(H$\delta$-strong, or H$\delta$S galaxies) represent a short-lived but
potentially important phase in galaxy evolution.  Such lines indicate a
stellar population dominated by A-stars which are normally either
absent or overwhelmed by the much brighter OB stars.  There is likely a
variety of physical mechanisms that can lead to such a stellar
population (e.g. Dressler et al.\ 1982; Couch et al.\ 1987; Poggianti
et al.\ 1999; Balogh et al.\ 1999), but most invoke a major
transformation from one galaxy type to another.  Although such galaxies
are very rare in the local Universe, their short lifetime means they
could potentially represent an important phase in the evolution of most
normal galaxies.

While high resolution imaging has demonstrated that most bright, nearby
H$\delta$S galaxies are spheroidal, often with signs of interaction
(Yang et al.\ 2004; Couch et al.\ 2004; Balogh et al.\ 2004), the
dynamics have been extremely difficult to observe and understand.
Norton et al. (2001) have obtained longslit spectra of galaxies in the
Zabludoff et al. (1996) sample, and find evidence that most of the
galaxies are in the process of transforming from
rotationally-supported, gas-rich galaxies to pressure-supported,
gas-poor galaxies.

However, longslit spectroscopy mixes spatial and spectral resolution,
and a better understanding of the dynamics can be obtained from
integral field spectroscopy.  This allows us to identify the physical
locations of the gas, old (K) stars, young OB-stars and (the
short-lived) massive A-stars, and to decouple their dynamics from their
spatial distribution.  In this paper we demonstrate the feasibility of
using an integral field unit (IFU) to study the dynamics of H$\delta$S
galaxies.  We have selected one field galaxy from the Sloan Digital Sky
Survey (SDSS) and used the GMOS-South IFU in Nod \& Shuffle mode to
study the rest-frame optical spectra.  Using this technique we can
investigate the spatial distribution of star formation through the
[O{\sc ii}] emission and the distribution of the young A-stars through
the (much stronger) stronger H$\delta$ and H$\gamma$ absorption lines.

In \S2 we present the data reduction and analysis.  The results are
presented in \S3.  Finally we summarise our results and present the
implications in \S4.  We use a cosmology with $\Omega_{m}$=0.3,
$\Lambda=$0.7 and a Hubble constant of 70$\kms$Mpc$^{-1}$. In this
cosmology, 2\,kpc subtends 1$''$ on the sky at $z=0.10$.

\section{Observations and Data Reduction}

The target is selected from the SDSS DR1 (Abazajain et al.\ 2001), and
was identified as a H$\delta$S galaxy by Goto et al.\ (2003).  This
field galaxy has a redshift $z=0.1055$, with J2000 coordinates RA:
10:13:45.39, Dec: +01:16:13.66.  Absorption lines in the SDSS spectra
were fit with a double--Gaussian profile model to account for narrower
emission filling within the absorption line, and all galaxies with
rest-frame equivalent widths (EWs) $>4$\AA\ were identified as
H$\delta$S.  For this pilot study, we chose a relatively bright
example, with $r=16.3$ mag (which corresponds to $\sim1/3L_{*}$).
SDSSJ101345.39 has strong absorption lines and weak (but
non-negligible) emission lines (Fig.~2).  Post-starburst galaxies with
strong H$\delta$ absorption, but with non-negligible [O{\sc ii}] or
H$\alpha$ emission lines are usually referred to as e(a) galaxies.
However as Balogh et al.\ (2004) show, e(a) galaxies with W$_{\rm
  o}$([O{\sc ii}])$<$10\AA\ and W$_{\rm o}$(H$\alpha$)$<$10\AA\ appear
to have the same properties of strict k+a galaxies (which are
H$\delta$S but have no detectable [O{\sc ii}] or H$\alpha$ emission).
k+a galaxies are predominantly bulge dominated with little sign of
spiral structure and signatures of recent (substantial) bursts of
star-formation (Balogh et al.\ 2004).  In this study, we have selected
one galaxy which falls in this {\it "k+a plus weak emission"} category
in order to compare and contrast the dynamics and spatial distribution
of the gas with that of the young stellar component.

%
%
\vspace{0.5cm}
\begin{inlinefigure}
  \vspace*{0.5cm}
  \centerline{\psfig{file=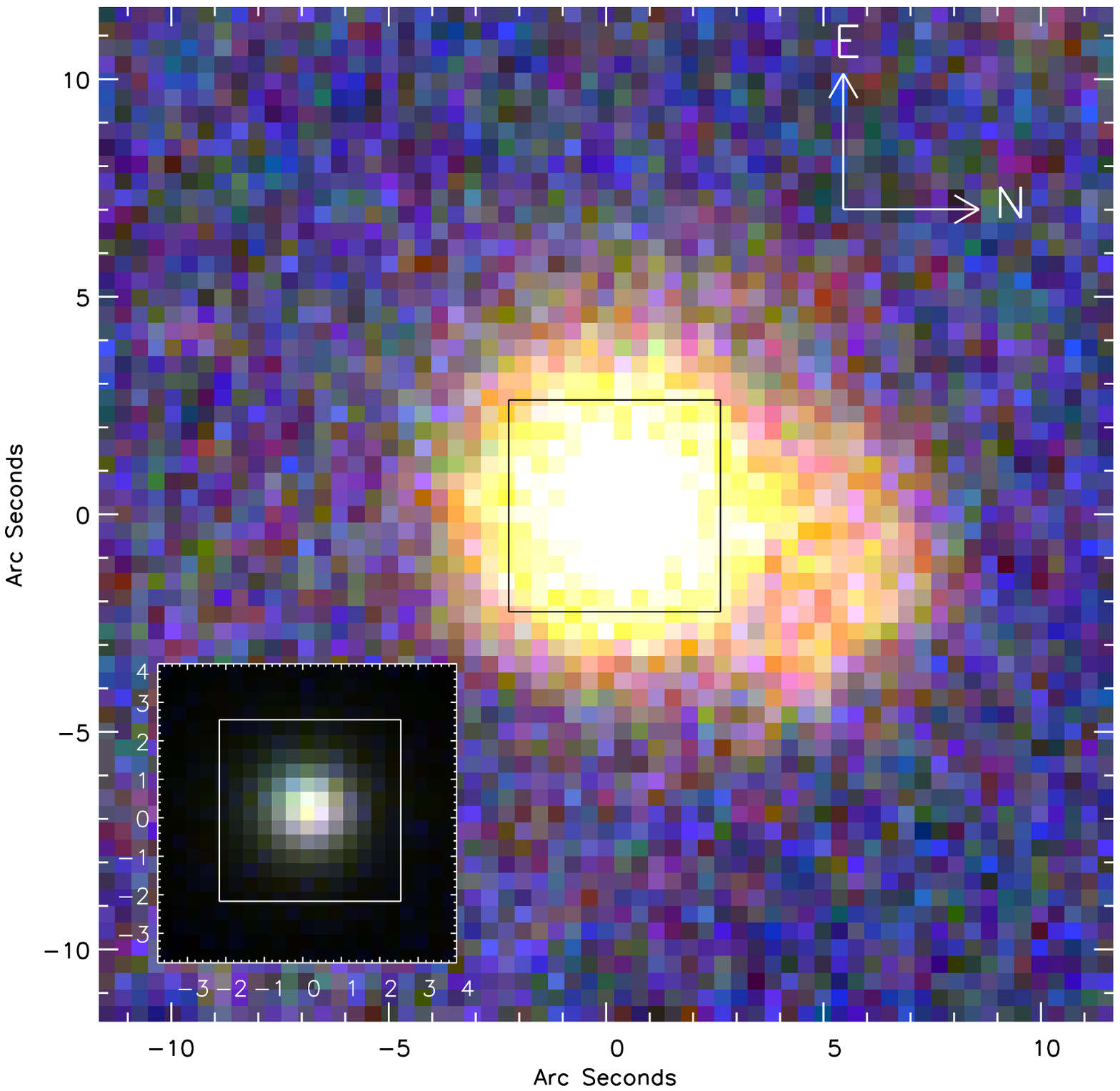,width=3.0in,angle=0}}
  \figcaption{True colour {\it gri} image of SDSSJ\,101345.39 from the
    SDSS imaging. Both panels show the GMOS-IFU field of view.  The
    main panel has been thresholded to emphasise the low surface
    brightness material outside the nucleus as well as the extension of
    material to the north, suggesting SDSSJ\,101345.39 may have
    recently undergone an interaction.  The inset is the same image but
    scales to emphasise the full range of surface brightness and shows
    that most of the stars are located inside the GMOS-IFU field of
    view.  We have rotated this image to agree with the IFU
    observation, and so North is right and East is up.}
\end{inlinefigure}

\subsection {Imaging}

Using the SDSS {\it gri}-band imaging we construct a true colour image
of SDSSJ\,101345.39.  Each image has an exposure time of 60 seconds and
was obtained in $\sim$1.2$''$ seeing.  The final images have a plate
scale of 0.4$''$/pixel.  The high surface brightness material in
SDSSJ\,101345.39 looks like a bulge dominated galaxy, however the lower
surface brightness structure has a disturbed morphology (Fig.~1) of
diffuse material, distributed asymmetrically toward the north, which
may be indicative of a tidal interaction or merger.  The GMOS-IFU field
of view covers the central nucleus (where most of the stars are
located), but the much lower surface brightness material is located
outside the IFU field and extends over a $\sim10"$ radius.

\subsection {SDSS Spectroscopy}
The SDSS spectroscopy of SDSSJ\,101345.39 shows strong
H$\gamma$\,$\lambda$4340.5\AA\ and H$\delta$\,$\lambda$4101.7\AA\ 
absorption lines, as well as emission lines.  From the original SDSS
spectra, we measure W$_{\rm o}$(H$\delta$)=7.5\AA\ in absorption, and
W$_{\rm o}$(O[{\sc ii}])=4.1\AA\ in emission.  The galaxy also shows
H$\alpha$\,$\lambda$6562.8\AA\ and [N{\sc ii}]\,$\lambda$6583.0\AA\ 
emission; although the measured equivalent width of H$\alpha$ is
7.9\AA, this is compromised by the strong underlying Balmer absorption.
Taking into account the underlying absorption, the emission line ratios
may be indicative of weak AGN activity in the galaxy (although as we
show in \S3, most of the emission is from a resolved component, and we
therefore attach a cautionary note that, from our current data, the
evidence for an AGN is tentative).

%
%
\begin{inlinefigure}
  \vspace*{0.5cm}
  \centerline{\psfig{file=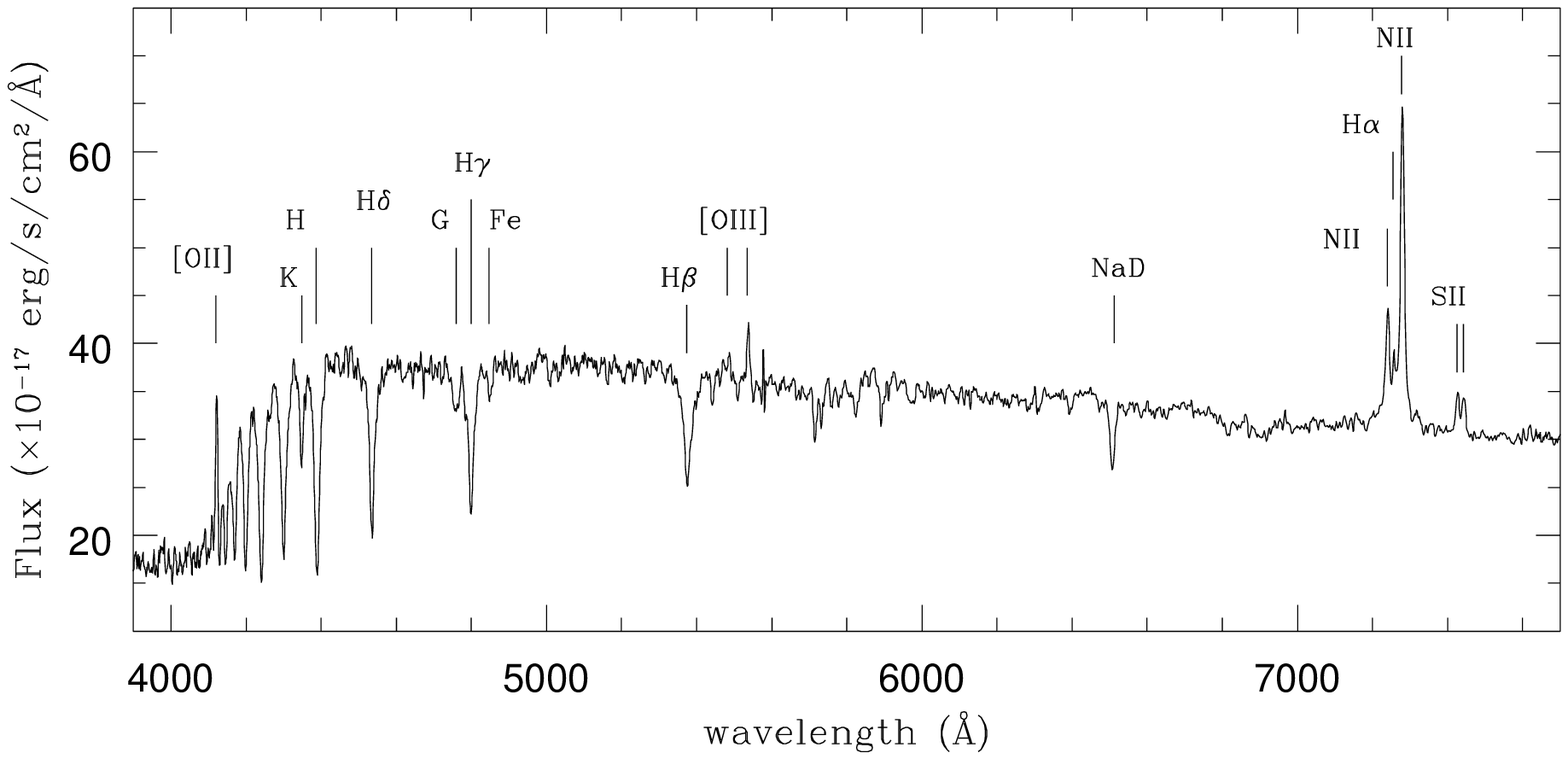,width=3.5in,angle=0}}
  \figcaption{SDSS spectra of SDSSJ\,101345.39.  This spectrum has a
    lower resolution than the GMOS spectrum, but shows the [O{\sc ii}]
    emission line as well as the much stronger H$\gamma$ and H$\delta$
    absorption lines.  The SDSS spectra also covers the H$\alpha$
    emission line.}
\end{inlinefigure}

%
%
\begin{figure*}
  \centerline{\psfig{file=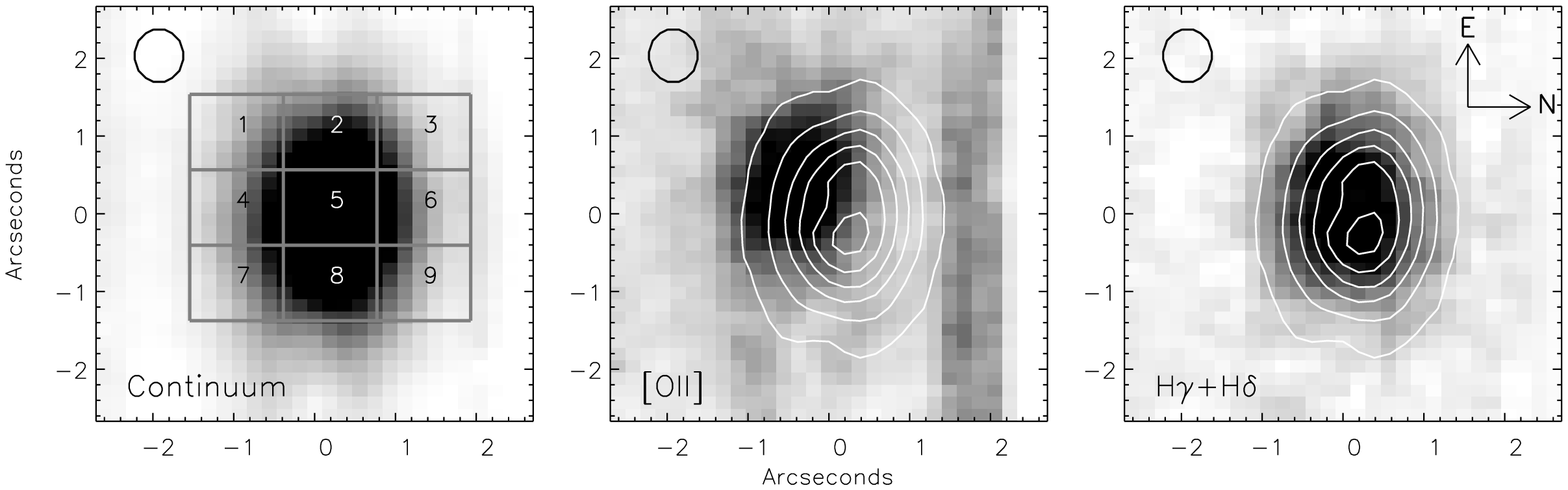,width=7in,angle=90}}
  \caption{Left: Continuum image generated from the datacube between
    4900--5400\AA; Middle: continuum subtracted two dimensional [O{\sc
      ii}] emission line map with the continuum from the left panel
    overlaid as contours.  Right: continuum subtracted two dimensional
    (inverted) H$\gamma$+H$\delta$ absorption line map with the
    continuum contours from the left panel overlaid.  The circles in
    the top left hand corners of each panel represent the size of the
    seeing disk.}
\end{figure*}

\subsection {GMOS Spectroscopic Imaging}

New observations of SDSSJ\,101345.39 were taken with the GMOS-South IFU
in Nod \& Shuffle mode on 2004 February 28$^{\rm th}$ U.T.\ during
science verification time for a total of 3 hours in $0.8''$ seeing and
photometric conditions\footnotemark.  Using the Nod \& Shuffle mode we
chopped away from the target by $30''$ every 30 seconds.  In this
configuration the IFU uses a fiber fed system to reformat the
$5''\times5''$ field onto two long slits.  Using the $B$-band filter in
conjunction with the B600 grating results in two tiers of spectra being
recorded.  The spectral resolution in this configuration is
$\lambda/\Delta\lambda\sim$1700.  The [O{\sc
  ii}]($\lambda\lambda$3726.1,3728.8\AA) emission line and H$\gamma$
and H$\delta$ absorption features all fall in regions of low sky
emission and absorption.

\footnotetext{Based on observations obtained at the Gemini Observatory,
  which is operated by the Association of Universities for Research in
  Astronomy, Inc., under a cooperative agreement with the NSF on behalf
  of the Gemini partnership: the National Science Foundation (United
  States), the Particle Physics and Astronomy Research Council (United
  Kingdom), the National Research Council (Canada), CONICYT (Chile),
  the Australian Research Council (Australia), CNPq (Brazil) and
  CONICET (Argentina).}

In the Nod \& Shuffle mode, the object and background regions are
observed alternately through the same fibres by nodding the telescope.
In between each observation the charge is shuffled on the CCD by a
number of rows corresponding to the the centre-to-centre spacing of the
blocks of 50 fibres into which each slit is divided. For the Nod \&
Shuffle mode, each alternate block is masked off so that it receives no
light from the sky but acts simply as an image store. The slit to field
mapping was arranged so that the resultant half of the total object and
background field (7$"\times$5$"$ + 3.5$"\times$5$"$) formed a
contiguous sub-field of 5$"\times$5$"$.  Since each alternate block is
masked, each exposure can be stored in different regions of the CCD
without contamination from any other region of the sky. The sequence of
object and background exposures can be repeated as often as desired
with the photoelectrons from each exposure being stored in their own
unique regions of the detector. At the end of the sequence, the CCD is
read, incurring a read-noise penalty only once (see Glazebrook \&
Bland-Hawthorn 2001) for further details of this general approach).
For each fiber (and corresponding spectrum), we identify the
corresponding sky spectrum in the shuffled position and subtract them
to achieve Poisson--limited sky subtraction.

%
%
\begin{inlinefigure}
  \vspace*{0.5cm} \centerline{\psfig{file=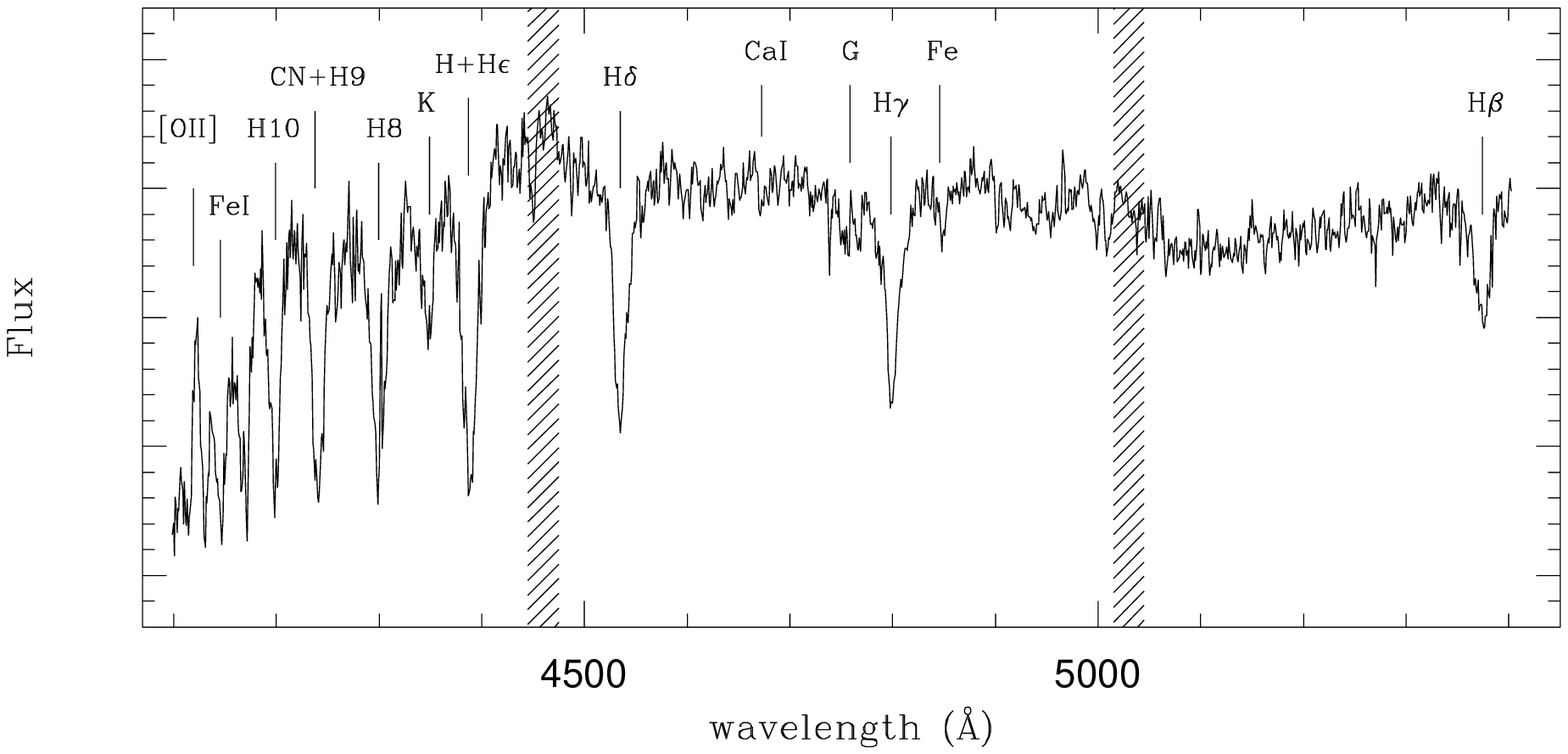,width=3.8in,angle=0}}
  \figcaption{The reduced, sky-subtracted spectrum of SDSS\,J101345.39,
    generated by collapsing the central two square arcseconds of the
    galaxy.  We have marked the strongest emission and absorption
    features.  The hashed regions show the positions of the GMOS chip
    gaps.}
\end{inlinefigure}

The GMOS data reduction pipeline was used to extract and wavelength
calibrate the spectra of each IFU element. The variations in
fiber-to-fiber response were removed using twilight flat-fields and the
wavelength calibration was achieved using a CuAr arc lamp.  The
wavelength coverage of the final data is 4080--5400\AA.  No flux
standards were observed, as the spectral features of interest are
narrow enough that the standard sensitivity is sufficient.

Since the datacube has a large wavelength coverage we correct for the
parallactic angle by modelling a 2-hour observation running from -1 to
+1 hour angle and corresponding minimum and maximum airmasses of 1.17
and 1.21.  The average atmospheric dispersion between 400nm and 540nm
is 0.67$"$ at an angle roughly North--South.  After building the
datacube, we use {\sc idl} to model and correct this aberration using a
linear interpolation at each slice of the datacube along the wavelength
axis.

We show the reduced, sky-subtracted spectrum, collapsed over the
central two arcseconds of the galaxy in Fig.~4 and identify the
strongest features.

%
%
\begin{figure*}[tbh]
  \centerline{\psfig{file=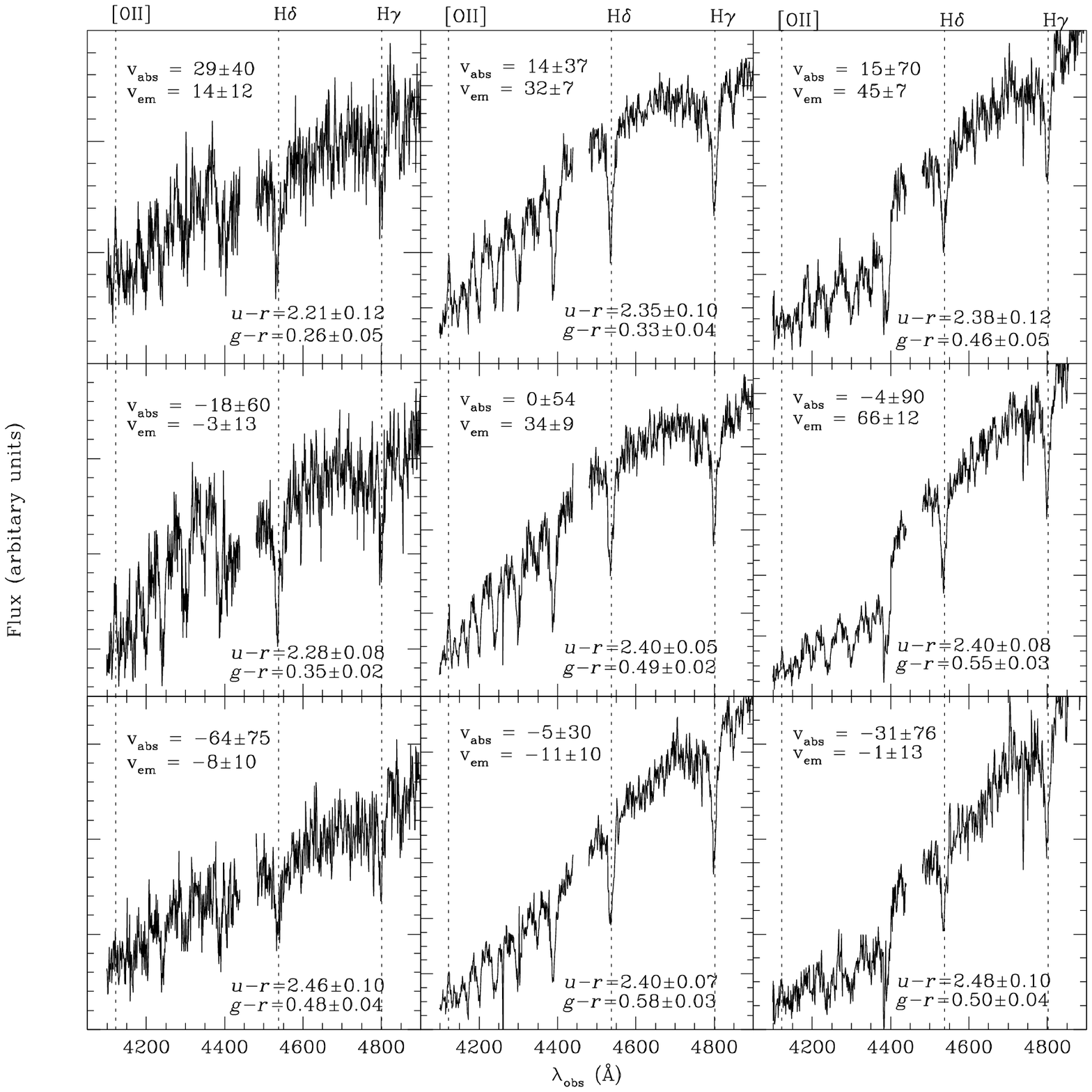,width=7in,angle=0}}
\caption{Spectra from nine spatial bins from the datacube split up 
  according to the labels in Fig.~3.  The dashed lines show the
  positions of the [O{\sc ii}], H$\gamma$ and H$\delta$
  emission/absorption lines for a fixed redshift of $z=0.1055$. The
  spectra which fall in the GMOS chip gaps have been masked out.  We
  also state the average [O{\sc ii}] emission line velocity, (v$_{em}$)
  found by fitting the [O{\sc ii}] emission line, as well as the
  absorption line velocity from the A- stars (v$_{abs}$) found by cross
  correlating each spectrum with an A2 stellar template (all velocities
  are in $\kms$).  The zero-point in the velocity is defined to be the
  rest-frame of the A-stars in the middle panel.  The change in the
  shape of the continuum which give rise to the colour gradients in
  Fig.~1 is reflected in the spectra.}
\end{figure*}

\section{Results}

\subsection{Spatial Light Distribution}

To investigate the spatial distribution of the gas and stars in the
galaxy we begin by extracting narrow-band slices from the datacube
around the emission and absorption lines.  We fit and subtract the
continuum around the line of interest using a 3--$\sigma$ clip to be
sure that neighbouring emission and absorption lines are omitted.  We
then extract the narrow-band images from the datacube by collapsing
each spectrum over the feature of interest.  We also extract the
continuum regions by median filtering each spectral pixel in the
datacube between 4900--5300\AA.

In Fig.~3 we show the spatial distribution of the relatively older
stars (as traced by the continuum light between 4900\AA\ and 5300\AA),
the gas (as traced through the [O{\sc ii}] emission line) and the
young, massive stars (traced by the H$\gamma$ and H$\delta$ absorption
lines).  Fig.~3 shows that, while both the [O{\sc ii}] emission and
Balmer absorption lines are extended, they do not have the same spatial
distribution.  In particular, the centroid of the [O{\sc ii}] emission
is offset $\sim 2$ kpc to the SE from the older stars and the
H$\delta$ and H$\gamma$ absorption lines.

This difference is shown in Fig.~5, where we split the datacube into
nine spatial bins.  It is clear that the shape of the continuum is not
constant across the galaxy.  This is confirmed by comparing with the
colour gradients in the SDSS imaging. By binning the imaging data into
$3\times3$ pixels (1.2$"\times1.2"$ bins, matching the same spatial
bins as in Fig.~5) we investigate the colour gradients across the
galaxy (which can be seen in the inset panel of Fig.~1). In each panel
of Fig.~5 we show the corresponding {\it g-r} and {\it u-r} colours
determined from the SDSS imaging data.  Both the spectroscopic and
imaging data show the colour gradient across the galaxy with panels
[1,2\&4],[3,5\&7] and [6,8\&9] having average {\it g-r} colours of
$0.31\pm0.06$,$0.48\pm0.06$ and $0.54\pm0.07$ and {\it u-r} colours of
$2.28\pm0.10$, $2.41\pm0.10$ and $2.42\pm0.09$ respectively confirming
that changing shape of the continuum seen in the IFU data is real and
not simply an artifact of the observations.

The spatial distribution of the young OB stars, as traced by the
rest-frame UV continuum, is consistent with that of the [O{\sc ii}]
emission, and offset by $\gsim$2\,kpc in projection from the older
stellar population (indicated also by the colour gradient in the
imaging).  The colour gradient is in approximately the same direction
as the surface brightness asymmetry, and suggests that the galaxy
consists of at least two components, possibly as the result of a recent
merger.

\subsection{Emission Line Dynamics}

To investigate the dynamics in more detail we return to the datacube
and fit the emission and absorption lines on a pixel-by-pixel basis.
The [O{\sc ii}] emission line doublet and underlying (rest frame UV)
continuum was fitted using a $\chi^2$ minimisation procedure.  The
spectra were averaged over a $3 \times 3$ spatial pixel region, except
where the signal was too low to give a significant detection of the
line, in which case the smoothing area was increased to $4 \times 4$
pixels. In regions where this averaging process still failed to give an
adequate $\chi^2$ (i.e. the inclusion of an emission line component
does not improve the fit), no fit was made.  We required a minimum
$\chi^2$ of 25 (S/N of 5) to detect the line, and allow the signal to
drop by a $\chi^2$ of 9 to calculate the error in the velocity.  This
corresponds to a formal 3$\sigma$ error.  In Fig.~6 we show the [O{\sc
  ii}] velocity structure.  This shows a velocity shear of
$100\pm20\kms$ across $\sim$6\,kpc in projection .  The velocity field
of the gas does not resemble that of a disk.  However, if one of the
progenitors was a gas rich disk, then the observed shear may be arise
due residual motion of a gas disk following a merger.  Assuming this is
the case, we estimate that the mass of the gas-rich progenitor was
($M=v^{2}r/G$) $\sim$1.2$\times$10$^{10}$$\Msol$.  This mass estimate
should be viewed as a lower limit to the progenitors mass since we may
be seeing only a residual part of the motion that the disk had before
the encounter.

%
%
\begin{inlinefigure}
  \vspace*{0.5cm}
  \centerline{\psfig{file=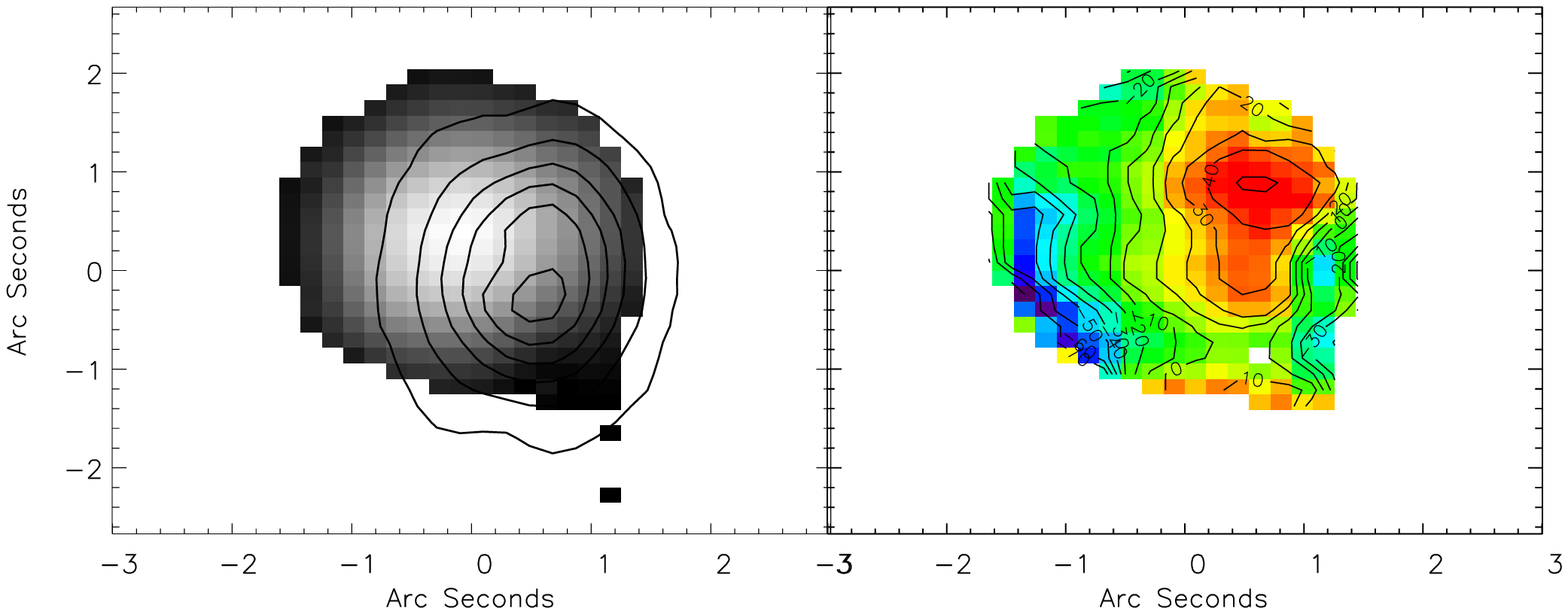,width=4.0in,angle=90,height=2.0in}}
  \figcaption{Left: The UV continuum intensity (colour scale) with the
    continuum from the older stellar population overlaid as contours.
    Right: Two dimensional velocity field of the galaxy derived from
    the [O{\sc ii}] emission line.  The galaxy shows a shear across the
    galaxy with peak-to-peak velocity difference of $100\pm20\kms$.}
\end{inlinefigure}

\subsection{A-type stars} \label{sec:astars}

To investigate the EWs of the short-lived A- stars we turn to the
H$\gamma$ and H$\delta$ absorption lines.  Using a similar procedure as
used to fit the [O{\sc ii}] emission features we simultaneously fit the
H$\gamma$ and H$\delta$ absorption features (with a fixed separation
but variable intensity and width) and show the resulting two
dimensional distribution of EWs in Fig.~7.  We estimate the 1--$\sigma$
uncertainties by perturbing all of the parameters of the best fit such
that $\Delta\chi^{2}=1$.  From the distribution of EWs it is clear that
the A- star population is most prominent at the same location as the
old stellar population, but also has a large contribution near the peak
in the [O{\sc ii}] distribution.  Whilst the two peaks in the
distribution of EWs in the galaxy are statistically significant, we
note that systematic uncertainties such as emission line filling of the
H$\gamma$ and H$\delta$ absorption lines may cause the EW to appear
artificially low.  We therefore attach little significance to the
`double peaked' distribution of EWs, but rather note that the
distribution of A-stars in the galaxy is widespread.  This is
particularly important for understanding the formation of the A-star
population since it suggests that the A-stars are not formed from gas
funnelled into the nucleus but rather from material which is much more
widespread.

\begin{inlinefigure}
  \vspace*{0.5cm}
  \centerline{\psfig{file=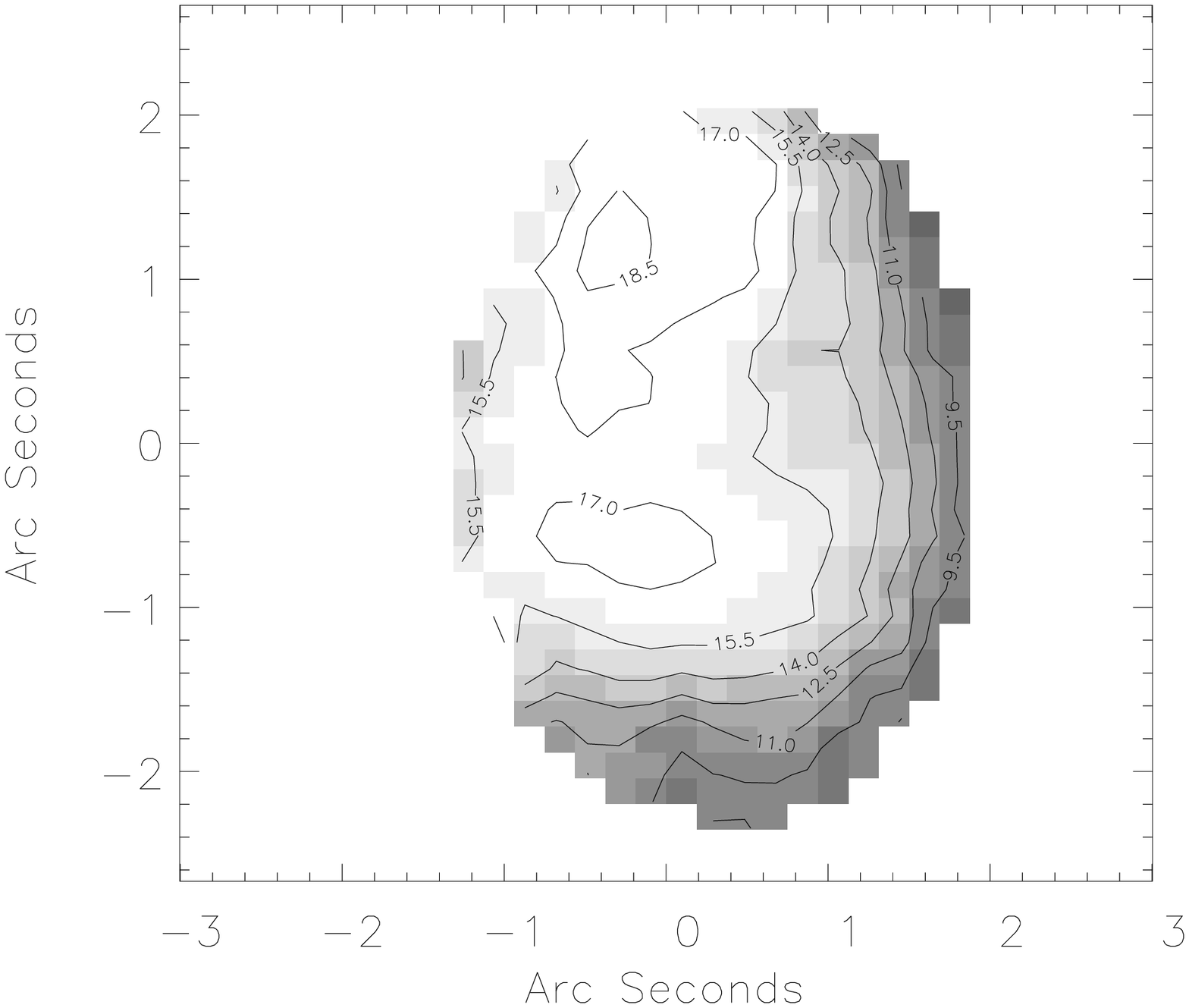,width=3.3in,angle=90}}
  \figcaption{Two dimensional map of the equivalent widths of the
    H$\delta$+H$\gamma$ absorption lines. This shows two peaks in the
    EWs, one at the peak location of the old stellar population, and
    one at the peak location of the younger OB stellar population.  The
    contours are spaced by 1--$\sigma$ found by varying the fit such
    that $\Delta\chi^{2}=1$.}
\end{inlinefigure}

To obtain the velocity structure of the A-type stars, the spectra from
the 9 spatial bins in Fig.~5 are cross-correlated with a template star
of type A2V. The absorption line velocities in individual panels are
shown in Fig.~5, with the zero-point defined such that the A-stars in
the middle panel have zero velocity.  The A-stars velocity field
appears to have a $\sim40\kms$ shear in the same direction of the gas,
although the lines have a greater intrinsic width and therefore the
large error bars preclude any stronger conclusion.

We can obtain an estimate of the age of the A-star population by
comparing the average galaxy spectrum with the spectral library of
Jacoby et al.\ (1984). The strength of the K-line at $\sim 3932$\AA\ is
especially sensitive to the spectral type of the early A-type stars
(Rose 1985).  Based on the relative strength of the K-line and the
neighbouring H+H$\epsilon$ line at $\sim 3970$\AA\, the galaxy spectrum
best resembles that of an A2$\pm 1$ star (hence the choice of the A2
stellar template for cross correlation above). Taking the effective
temperature of an A2 dwarf to be 9100 K (di Benedetto 1998), and
interpolating the theoretical isochrones of Bertelli et al.\ (1994) we
arrive at an estimate of 0.44 Gyr for a MS turnoff at A2. Likewise, the
ages for A1 and A3 dwarfs are estimated to be 0.39 and 0.47 Gyr
respectively although stellar mixes and metallicity makes this estimate
uncertain.

\section {Discussion and Conclusions}

We have selected one field galaxy from the SDSS which shows strong
Balmer absorption lines (which are indicative of a strong A- star
population), but has only a small ongoing star-formation rate.  In this
pilot study, we have demonstrated the ability of IFU observations to
disentangle the spatial and velocity distributions of the stellar and
gaseous components through the measurement of both emission and
absorption features.

The main results are summarised as follows:

\begin{enumerate}
\item The SDSS images show a bright nucleus and more diffuse material
  distributed asymmetrically toward the north (Fig.~1). The disturbed
  morphology of the field galaxy SDSSJ\,101345.39 is suggestive of an
  interaction (or merger) between two galaxies.
\item The continuum from the older stellar population is spatially
  offset by $\sim$2\,kpc from the young, hot OB stars (traced from the
  rest-frame UV continuum) and is easily seen from the changing shape
  of the continuum across the spatial domain of the datacube in the IFU
  observation (Fig.~5).  This colour gradient is also reflected by the
  strong colour gradient seen in the SDSS imaging.
\item By extracting narrow-band slices of the emission and absorption
  lines in the the GMOS IFU datacube, we find that the gas (traced
  through the [O{\sc ii}] emission line) is spatially extended with an
  elongated morphology, approximately coincident with the young, hot,
  OB star population (Fig.~3).
\item The $H\gamma$ and $H\delta$ absorption lines are widely
  distributed (Fig.~3).  The peaks of these absorption lines are
  spatially co-incident with the older stellar population and the
  centre of the gas.  The A-star population are not confined to the
  nuclear regions.  Their spatial distribution is different from that
  of the [O{\sc ii}] emission, but the two appear related.
\item The gas has a velocity shear of $100\pm20\kms$ across 6\,kpc in
  projection (Fig.~6), but the velocity field does not resemble that of
  a disk.  However, assuming that the velocity shear arises due to
  residual motion of a gas disk which has been disturbed as the result
  of an interaction, we estimate that the gas rich progenitor had a
  dynamical mass of at least $\sim1.2\times10^{10}\Msol$.  This mass
  estimate should be viewed as a lower limit since some of the angular
  momentum of the gas-rich progenitor may have been lost during the
  interaction.
\item The A-stars do not show the same amplitude of velocity motion as
  the gas, however, there is evidence of a 40$\kms$ shear in
  approximately the same direction as the velocity motion of the gas
  (although the large error bars preclude any strong conclusions).
\end{enumerate}

Our observations suggest that the activity in SDSS\,J102145.39 may be
the result of a strong tidal interaction between a passive (possibly
elliptical) galaxy (giving rise to the old stellar population), and a
gas-rich, star-forming (spiral or irregular?) galaxy (giving rise to
the spatially offset of hot stars and nebular emission).  The
interaction between the two components is likely to have been
responsible for the production of the A-star population in a burst of
star formation.
  
By comparing the galaxy spectrum with spectral libraries and stellar
population models (Jocoby et al.\ 1984, Vazdekis \& Arimoto 1999) we
estimate an burst age of $\sim$0.5Gyr although we caution that
systematic uncertainties such as hidden emission, differing
metallicities and pollution of K- type stars make this estimate
uncertain.  Unfortunately the wavelength coverage of our GMOS data do
not extend to Mg{\sc ii} or H$\alpha$, making it difficult to constrain
the underlying old population or dust absorption. Future observations
at these wavelengths would allow us to trace the dynamics and
metallicities of the old stars and further constrain the burst age, as
well as locate the position of any possible AGN (which may be important
for understanding black hole growth in these galaxies).

We can estimate the mass of the underlying old population from the
$K$-band magnitude (K=13.2+/-0.001, Balogh et al.\ 2004) and assuming a
canonical value of 1.0 for the stellar mass-to-light ratio (Bruzual \&
Charlot 2003).  We estimate that the total stellar mass is
$\sim2\times10^{11}\Msol$. It is interesting to contrast this with the
mass of the young A-star population, and the current star formation
rate. If we assume a Salpeter IMF (Salpeter 1955), we can estimate the
mass of the burst population by comparing the equivalent width of
H$\delta$ (7.5\AA) with the models of Shioya et al.\ (2004).  In order
to produce absorption of this strength, their models require a burst
population of at least $\sim 10\%$ by mass (ie.,
$\sim2\times10^{10}\Msol$). We note that this agrees well with the
lower limit on the dynamical mass that we estimated from the velocity
shear in the [O{\sc ii}] emission (we note that it is difficult to
place an upper limit on the mass of the gas rich progenitor). If the
burst mass fraction was skewed towards higher mass stars, the burst
mass could be a factor $\sim 3$ lower. The current residual star
formation rate inferred from the [O{\sc ii}] flux is
$\sim0.5\pm0.1\,{\rm M}_{\odot}{\rm yr}^{-1}$ (Kennicutt 1998), so that
in 0.5Gyr, a mass of only $\sim2\times10^{8}\Msol$ would be converted
into stars.  Thus it seems that the residual star formation is at least
1/100$^{th}$ of the rate during the burst, while the mass involved in
the burst is about 1/10$^{th}$ of the mass of the old stellar
population.

The spectra, morphological information and kinematics of
SDSSJ\,121345.39 appears very similar to local field E+A galaxy
NGC\,2865 (Hau, Carter and Balcells 1999) in which a 0.4--1.2Gyr old
burst has occurred due to the merger of a gas-rich (Sb or Sc) spiral
and an elliptical galaxy.  A quantitative discussion of the
star-formation truncation mechanism will have to wait until we have a
statistically significant sample, but we can outline the possibilities
which cause this process.  Competing models for E+A galaxy formation
predict very different distributions of stellar populations following
the interaction.  In SPH simulations for mergers between spiral and
elliptical galaxies the gas tends to sink to the galaxy centre on
timescales much less than the starburst duration, and therefore suggest
that the gas and A-stars should be centrally concentrated and
segregated away from the older stellar population (e.g.,\ Hernquist \&
Weil 1992) (although these models do not take into account the hot ISM
in the elliptical).  However, other models suggest that the gas in the
progenitor should be clumpy (in molecular clouds) rather than
distributed as a smooth medium, and the result of the merger is to
scatter these star forming regions across the galaxy
(compressing/squeezing the gas and causing the A-star starburst as the
interaction occurs).  As the gas clouds scatter either in the spiral
arms of the progenitor or in the new ICM, they evaporate and the star
formation is truncated as the density of the clouds falls (eg.Mahos \&
Hernquist, 1996, Kojima \& Noguchi 1997).  In our pilot study we have
shown that the A-stars are widely distributed across the galaxy, which
coupled with the spatially extended morphology of the gas suggests a
clear preference for the latter model.

Although these observations show a clear preference for the {\it
  squeezed cloud} model it is not yet clear that this is true for the
H$\delta$S galaxy population as a whole.  The next step is to generate
a statistically useful sample of such galaxies, in a variety of
environments, to be able to generalise our conclusions and investigate
the star-formation rate of the present day Universe.  This is important
for constraining current models for galaxy formation in which these
H$\delta$S galaxies may represent an important phase in the evolution
of todays local luminous spheroidal galaxies.

\acknowledgements We are very grateful to the Gemini Staff for
accepting this programme for Science Verification Time.  We would like
to thank Chris Simpson and Ian Smail for useful discussions.  AMS
acknowledges financial support from PPARC. RGB holds a PPARC senior
fellowship.

\end{document}